# Tropical Cyclone Intensity Evolution Modeled as a Dependent Hidden Markov Process

Renzhi Jing (corresponding author, rjing@princeton.edu) and

Ning Lin (nlin@princeton.edu)

Department of Civil and Environmental Engineering, Princeton University

Abstract

A hidden Markov model is developed to simulate tropical cyclone intensity evolution dependent on the surrounding large-scale environment. The model considers three unobserved (hidden) discrete states of intensification and associates each state with a probability distribution of intensity change. The storm's transit from one state to another is described as a Markov chain. Both the intensity change and state transit components of the model are dependent on environmental variables including potential intensity, vertical wind shear, relative humidity, and ocean feedback. This Markov environment-dependent hurricane intensity model (MeHiM) is used to simulate the evolution of storm intensity along the storm track over the ocean, and a simple decay model is added to estimate the intensity change when the storm moves over land. Data for the North Atlantic (NA) basin from 1979 – 2014 (555 storms) are used for model development and evaluation. Probability distributions of 6-h and 24-h intensity change, lifetime maximum intensity, and landfall intensity based on model simulations and observations compare well. Although the MeHiM is still limited in fully describing rapid intensification, it shows a significant improvement over previous statistical models (e.g., linear, nonlinear, and finite mixture models).

1. Introduction

Tropical cyclones (TCs) are one of the most destructive natural phenomena on Earth and are the leading cause of great social and economic losses. Accurately assessing TC-related risk is of great importance. However, while TC intensity plays a large role in the damage potential of a storm, TC intensity modeling is still far from satisfactory (Rappaport et al. 2009). This is mainly because the intensity evolution of TCs involves complicated dynamic and thermodynamic processes that have not been fully resolved (Camargo and Wing 2016, Walsh et al. 2016). Apart from internal variability (e.g., eyewall replacement cycles; Willoughby et al. 1982), the storm environment plays a key role in controlling the storm's intensity evolution. The pre-storm sea surface temperature (SST) has been considered as one of the most important environmental variables (Whitney et al. 1997), since TCs draw their energy from the underlying seawater. Emanuel et al. (2004) used an ocean-coupled model to show that other environmental factors, including maximum potential intensity (MPI) and vertical wind shear, also have great influences on storm intensity. Furthermore, the interaction between TC and the ocean can significantly influence TCs' intensity. The entraining of cold, deep water induced by storm's surface winds may cause strong cooling of SST and thus negatively affect a storm's intensification (Schade and Emanuel 1999; Lloyd and Vecchi 2011).

Under the cooperative work of inner-core, oceanic, and large-scale environmental processes, many TCs undergo rapid intensification (Kaplan and DeMaria 2003) during their lifecycles. Rapid intensification (RI) is the significant strengthening of a TC in a short time (specifically, intensity increases by 30 kts or more over 24 hours), and almost all historical category 4 and 5 hurricanes are RI storms (Kaplan and DeMaria 2003). Accurately predicting RI not only is important in real-time forecasting (Toepfer et al. 2010), but also plays an important role in reproducing TC intensity climatology, as the RI storms control the tail (including possibly a secondary peak) of the statistic distribution of a TC's maximum lifetime intensity (LMI) (Lee et al. 2016a). However, the prediction of RI is still a great challenge, and few operational models are skillful in predicting RI (Elsberry et al. 2007). To fill the gap, Kaplan and DeMaria (2003) defined the rapid intensification index (RII) as a probabilistic guidance. A revised RII is further developed to estimate the probability of RI occurrence over the succeeding 24-h for various RI thresholds and for different basins (Kaplan et al. 2010). Although the probabilistic RII forecast is generally skillful compared to the baseline climatology, the false alarm rate of RII is still high, especially for the North Atlantic Basin (Kaplan et al. 2010).

Despite the challenges, various statistical methods have been applied to model TC intensity change as a response to environmental forcing. One widely used model is the Statistical Hurricane Intensity Prediction Scheme [SHIPS; (DeMaria and Kaplan 1994; DeMaria and Kaplan 1999)], which estimates TC intensity change as a linear regression function of key environment variables (e.g., MPI, vertical wind shear, upper-level winds) and TC characteristics (e.g., current intensity, previous intensity change, and storm's translation speed). Lee et al. (2015) narrowed down the essential predictors in SHIPS into a smaller pool and developed a simpler linear regression model. Lee et al. (2016b) improved this model by adding a stochastic component to better capture the statistic distributions of TC intensification rate and LMI.

More recently, Lin et al. (2017) applied advanced statistical techniques to further explore the dependence of TC intensity change on the environmental variables, including the newly developed ventilation index (VI; Tang and Emanuel 2012). They found that applying the three component variables of VI (i.e., MPI, vertical wind shear, and entropy deficit) is statistically better than applying VI as a single variable, and the simpler relative humidity represents the entropy deficit relatively well. After identifying the most important environmental variables, Lin et al. (2017) developed linear, nonlinear, and mixture regression relationships between TC intensity change and these variables. In particular, they found that the statistical $R^2$ of their linear and nonlinear models is similar to that of SHIPS, but the statistical $R^2$ of their mixture model (involving multiple subgroup models) is significantly higher, indicating that the dependence of TC intensity change on the environment is nonhomogeneous. More interestingly, their mixture regression analysis identified three regimes/states of intensity change: static, normal, and extreme, reflecting a storm's slow, moderate, and rapid intensity change, respectively, during its lifecycle (Lin et al. 2017). However, applying the mixture model to predict TC intensity evolution requires adding a classification model to identify which regime the storm belongs to at each time step, which can induce large uncertainties/errors, especially for extremes related to RI (Lin et al. 2017).

Motivated by the findings of Lin et al. (2017), here we model TC intensity evolution as a dependent hidden Markov process. This new model is called the Markov Environment-dependent Hurricane Intensity Model (MeHiM). Like the mixture model, the MeHiM considers three unobserved (hidden) states of intensity change and associates a probability distribution of intensity change with each state. Unlike the (statistical) mixture model, which assumes that the state identity for each time step is determined separately, the (stochastic) MeHiM considers the storm's transit from one state to another as a Markov chain.

Both the intensity change and state transition components of the MeHiM are dependent on identified environmental variables. To keep the model relatively simple, we apply only four environmental variables (MPI, vertical wind shear, relative humidity, and an ocean feedback parameter) and two storm variables (current intensity and previous intensity change). Few of the developed ocean feedback parameters (e.g., Schade and Emanuel 1999, Vincent et al. 2012, Balaguru et al. 2015), which indicate ocean-storm interaction, have been evaluated in statistical modeling of TC intensity. Here we adopt the ocean feedback parameter of Schade and Emanuel (1999), which has also been used in a simplified dynamical model (Emanuel 2017), and test its impact on the overall performance of the MeHiM.

The MeHiM is developed to predict the intensity change of storms moving over the ocean. To simulate the intensity evolution over storm's entire lifecycle, we add a simple decay model to predict a storm's intensity change over land. Here we develop the model for the North Atlantic basin, though the same methodology can be applied to other basins or on the global scale. To evaluate the developed model, we apply it to simulate historical storms using historical tracks and environment taken from reanalysis datasets. The performance of the model is measured by its capability to reproduce the observed TC intensity metrics, including spatial distribution of intensity over the basin, probabilistic distributions of intensification rate, LMI, and landfall intensity, as well as temporal evolution of intensity for sample cases. We focus particularly on the model's capability to capture the extremes related to RI.

The MeHiM improves on the mixture model of Lin et al. (2017), because 1) it accounts for the (possibly large) temporal correlation of the states (e.g., slow, moderate, rapid) of intensity change over storm's lifecycle and 2) it integrates the mixture model of intensity change and the identification model of intensification state within a Markov framework, avoiding statistical inconsistency between the two components. It is interesting to ask whether these improvements in methodology will lead to significant improvements in the model performance, especially in capturing RI related extremes. To answer this question, we build the mixture model (with an added identification model) using the same probability distributions (i.e., Gaussian for intensity change and Logistic for state identification), environmental and storm variables, and observational datasets as in the MeHiM. In addition, we build the (Gaussian) linear regression model using the same variables and datasets to provide a baseline. The linear and mixture models are run in parallel with the MeHiM, and the simulation results are compared among the models and with the observations.

This study is organized as follows. Following this introduction, the development of the models (i.e., linear regression model, mixture regression model, MeHiM, and land decay model) is described in Section 2. Model simulation results are shown and compared in Section 3. Section 4 discusses the effects of adding the ocean feedback parameter and correctly identifying the extreme state of intensity change in the MeHiM. Section 5 summarizes the main findings of the study.

2. Model Development

Three statistical TC intensity models are developed to predict 6-hourly storm intensity change DV: the multiple linear regression model, mixture model (with an added state identification model), and MeHiM. We build these models on the same six variables: last step intensity change ($DV_p$), the current intensity (V), maximum potential intensity (MPI), vertical wind shear (SHR), relative humidity (RH), and an ocean predictor (OCN). Apart from these intensity models, which are used to simulate a storm's intensification over the ocean, we also add a simple land model to simulate the decay of the storm intensity over time when the storm moves on land.

All models are developed based on historical records from 1979 to 2014 in the North Atlantic basin. The TC dataset is taken from the IBTrACS WMO archive (Knapp et al. 2010). It includes 6-hourly latitude and longitude positions and 10-minute maximum sustained wind speeds at 10 m above the sea surface. The location data are used to calculate the translational velocity of the storm. The storm maximum wind intensity (V) is estimated by subtracting the surface background wind, estimated as a fraction of the storm translation velocity (Lin and Chavas 2012), from the observed maximum wind. The 6-hour intensity change DV (and $DV_P$) is then obtained from the time series of the estimated storm maximum wind intensity.

The atmospheric variables are derived from the ERA-Interim Reanalysis with a resolution of 0.75º x 0.75º produced by the European Centre for Medium-Range Weather Forecasts [ECMWF; (Dee et al. 2011)]. A storm's maximum potential intensity, MPI, is theoretically derived following Emanuel (1995) and Bister and Emanuel (1998; 2002). Similar to that in previous studies (e.g., DeMaria and Kaplan 1994, Lin et al. 2017), the deep layer vertical wind shear, SHR, is defined as the difference between the 850- and 200-hPa level winds, averaged over a 200-800 km annulus around the storm center. High-level relative humidity, RH, is computed as the averaged relative humidity in the layer between 300- and 500-hPa within a 500-800 km annulus around the storm center.

In addition to the atmospheric predictors, we also incorporate an oceanic variable (OCN), developed by Schade and Emanuel (1999) based on numerical modeling, to represent the ocean's negative impact on storm intensification. OCN is defined as

$$\text{OCN} = 1 - 0.87 e^{-z} \tag{1a}$$

where

$$z \equiv 0.01 \Gamma^{-0.4} h_m u_T \frac{\text{PI}}{V} \tag{1b}$$

where $u_T$ is the storm translation speed, $h_m$ is the ocean mixed layer depth, and $\Gamma$ is the thermal stratification below the ocean mixed layer. OCN is a number between 0 (strong weakening influence on the storm) and 1 (weak influence); it is closer to 0 when the storm is stronger, moving more slowly, with a shallower mixed layer depth, and/or with stronger thermal stratification. OCN is more sophisticated than other ocean variables such as the ocean temperature averaged over the mixing layer and the ocean heat content, which indicate only the amount of heat stored in the ocean and are incapable of representing the interaction between the storm and ocean. The ocean heat content is used in SHIPS, but the data are available only for a limited time period and over a limited part of the Atlantic basin (Mainelli et al. 2005); the ocean temperature averaged over the mixing layer was found to be an insignificant predictor of TC intensification (Lee et al. 2015). To estimate the variables in OCN, we take the ocean's salinity and potential temperature from the Ocean Reanalysis System 4 [ORAS4; (Balmaseda et al. 2013)]. The variable $h_m$ is computed following de Boyer Montégut et al. (2004) with a temperature difference criterion of $0.5°C$. The variable $\Gamma$ is estimated as the temperature gradient within 100 m below the mixed layer depth $h_m$.

The following four subsections describe the development of the three over-the-ocean intensity models and the land decay model with statistical fitting results. In all statistical analyses, variables are standardized so that all variables are on the same scale. To maximize the utilization of the data, the models are developed with all 555 TCs from 1979 to 2014. The models are evaluated in terms of their capability of capturing the statistical distributions of various intensity and intensification metrics and simulating realistic sample cases of intensity evolution, as discussed in the following sections.

### 2.1. Ordinary Least Squares (OLS)

For a multiple linear regression model with OLS, the probability distribution of the predictand $Y$ is Gaussian, as shown in Eq. (2), where $Y$ is the 6-h intensity change ($DV$), and $\boldsymbol{X}$ is the vector of predictors (i.e., $DV_p$, V, MPI, SHR, RH, and OCN).

$$Y \sim \mathcal{N}(\alpha + \boldsymbol{X}\boldsymbol{\beta}, \sigma) \tag{2}$$

The model coefficients $\alpha$ and $\boldsymbol{\beta}$ are estimated by minimizing the squared residual error (i.e., OLS). The estimated model coefficients as well as the standard deviation ($\sigma$) are shown in Table 1. The signs of the coefficients in front of all variables are as expected (based on previous studies; e.g., DeMaria and Kaplan 1999, Schade and Emanuel 1999, Lin et al. 2017), i.e., DV is positively (negatively) correlated to $DV_p$ (V); it is expected to increase when MPI and RH are larger, SHR is smaller, and/or ocean cooling is weaker (OCN larger). The storm's last-step intensification $DV_p$ is shown to be the strongest predictor, followed by the current intensity (V). Among the environmental variables, MPI and SHR are more important than RH and OCN. The intercept ($\alpha$) is nearly zero, which means that the expected value of 6-h intensity change DV is nearly 0 when all five predictors are at their means (i.e., standardized values being zeros).

## 2.2. Finite Mixture Regression (FMR)

The Finite Mixture Regression (FMR) extends the OLS by incorporating a heterogeneity structure to the model. In this model, each observation is considered to be a random sample generated by one of its k components (groups/clusters), while the identity of the generating component is not observed. In a Gaussian setting, each of the k components is a linear regression, and the probability distribution of the predictand Y is mixed Gaussian,

$$Y \sim \sum_{r=1}^{k} w_r \mathcal{N}(\alpha_r + X\boldsymbol{\beta}_r, \sigma_r) \tag{3}$$

where $w_r$ is the component mixing proportion/weight, $\alpha_r$ and $\boldsymbol{\beta}_r$ are regression coefficients, and $\sigma_r$ is the standard deviation, for each component $r$. These model parameters can be estimated from data using the maximum likelihood estimation method with the generalized expectation-maximization (EM) algorithm (McLachlan and Krishnan 2007). In addition to model parameters, the EM algorithm also estimates the probability that each observation was generated from each of the $k$ groups, providing an optimal group assignment for each observation.

Based on the same dataset as for the OLS, the estimated model parameters for FMR with three groups ($k$=3) are shown in Table 2(a). The obtained group weights and features are similar to those from the FMR analysis of Lin et al. (2017). Specifically, over 60% of the observations are optimized by the EM algorithm into Group 2, nearly 30% into Group 1, and only less than 10% into Group 3. Groups 1, 2, and 3 again feature "static," "normal," and "extreme" intensity change (DV), respectively, in the sense that, first, the mean of the intensity change as a response to the environment is modeled to be weak for Group 1, relatively strong for Group 2, and the strongest for Group 3 (as indicated by the magnitude of the model coefficients in front of the environmental variables); and, second, the model standard deviation (representing the width of the subgroup Gaussian distribution) increases significantly from Group 1, to Group 2, and to Group 3. As a result, the sample mean of DV increases moderately in absolute value from Group 1 (-0.07), to Group 2 (0.82), and to Group 3 (-3.86) and the sample standard deviation of DV increases substantially from Group 1 (0.47), to Group 2 (5.89), and to Group 3 (16.73), indicating that (regardless of the environmental conditions) small intensity change observations are assigned into Group 1, extreme intensification (and deintensification) into Group 3, and the rest into Group 2 (according to the optimal assignment based on probability). Such grouping features will be further discussed with the Markov model analysis. It is also noted that some unexpected signs of the model coefficients are obtained in the FMR modeling, e.g., negative response to MPI as a statistical compensation for the large negative response to SHR in Group 3, which will be shown to be improved in the Markov model.

The FMR is not a predictive model, and conducting simulations with FMR requires assigning group weights or membership for new observations. Similar to Lin et al. (2017), we add a multinomial logistic regression (Eq. 4) as a classification model to determine the group membership probabilities $\pi_r$ for the new observation $X$ (i.e., $DV_p$, V, MPI, SHR, RH, and OCN),

$$\pi_r(X) = \frac{\exp(\theta_r + X\gamma_r)}{\sum_{r=1}^{k} \exp(\theta_r + X\gamma_r)} \tag{4}$$

The multinomial regression model is fitted based on the FMR optimized group memberships, and the obtained model coefficients ($\theta_r, \gamma_r$) are shown in Table 2(b). We pick Group 3 as the baseline group (all coefficients are assigned to be 0). The linear combination of covariates for Group 1 (first row) describes the log odds of belonging to Group 1 vs. belonging to Group 3, and, similarly, the linear combination of covariates for Group 2 (second row) describes the log odds of belonging to Group 2 vs. belonging to Group 3. For example, a one-unit, or one standard deviation, increase in the variable $DV_p$ is associated with a decrease in the log odds in the amount of 0.359, which is equivalent to a decrease in the odds by approximately 30%. At the same time, an increase in the variable $DV_p$ is also associated with a decrease in the odds of belonging to Group 2 vs. belonging to Group 3 by 25%. Thus, with other covariates fixed, an increase in $DV_p$ will cause the new observation more likely to be assigned to Group 3.

To obtain a more general view of the obtained classification model, we estimate the group membership probabilities for three specific scenarios. When all variables of the new observation are at their medians (over all samples in the dataset), the probabilities that it belongs to Groups 1, 2, and 3 are 0.41, 0.58, and 0.01, respectively. When all variables of the new observation are at the 95$^{th}$ percentile favorable (according to the expected coefficient signs shown in Table 1) for larger DV, the probabilities that it belongs to Groups 1, 2, and 3 are 0.39, 0.60, and 0.01, respectively. When all variables of the new observation are at the 95$^{th}$ percentile unfavorable for larger DV, the probabilities that it belongs to Groups 1, 2, and 3 are 0.23, 0.74, and 0.03, respectively. The probabilities obtained in the three scenarios are relatively close, indicating that the model may be limited in capturing the change in the membership probability as a response to the change of the predictor variables. Moreover, the probability of belonging to the extreme group (Group 3) is always significantly smaller than the group weight from the optimization (~ 0.07; Table 2a). This limitation in capturing the extreme state by the FMR classification model was also noted by Lin et al. (2017). This limitation is reduced in the Markov model.

2.3. The Markov environment-dependent Hurricane Intensity Model (MeHiM)

The MeHiM is based on the idea of the hidden Markov model (HMM; Rabiner 1989, Cappé et al. 2006), an effective tool for modeling stochastic processes. In an HMM, the visible observations are distributed according to the underlying hidden states, where the states are discrete and are assumed to form a first-order Markov process. With a selected number ($k$) of states, an HMM can be fully determined by the following three sets of parameters:

1. Initial state probability $\pi_i$: the probability of starting from state $i$.
2. Transition probability $\pi_{ij}$: the probability of moving from state $i$ to $j$.
3. Response distribution $f_{Y_i}$: the distribution of the response variable in state $i$.

The HMM can be extended to the dependent HMM with the above model parameters dependent on covariates (Zucchini et al. 2016). Then, the likelihood of the response observation sequence $y_{1:T} = y_1 y_2 \ldots y_T$, given the covariates sequence $\bm{x}_{1:T} = \bm{x}_1 \bm{x}_2 \ldots \bm{x}_T$, is

$$L(y_{1:T}|\bm{x}_{1:T}) = \sum_{s_1=1}^{k} \pi_{s_1}(\bm{x}_1) f_{Y_{s_1}}(y_1|\bm{x}_1) \prod_{t=2}^{T} \pi_{s_{t-1} s_t}(\bm{x}_t) f_{Y_{s_t}}(y_t|\bm{x}_t) \tag{5}$$

where $s_t$ denotes the state at time $t$, and all three model components (i.e., initial state probability, transition probability, and response probability) depend on the covariates.

For the entire dataset, the total likelihood is the product of the individual likelihood functions for all time series.

The basic idea of a dependent HMM is applied in the MeHiM to model the sequence of TC intensity change (DV) over the storm's lifetime dependent on the selected storm and environmental covariates (i.e., $DV_p$, V, MPI, SHR, RH, and OCN). Motivated by the FMR analysis, we assume the storm in one of three hidden states ($k = 3$) at each time step. Given the state, the response distribution is assumed to be Gaussian, i.e.,

$$Y_i \sim \mathcal{N}(\alpha_i + \boldsymbol{X\beta}_i, \sigma_i) \tag{6}$$

dependent on the 6 predictor variables as in the OLS and FMR models.

Similar to the FMR analysis, we apply the multinomial logistic regression to model the transition probability that the storm moves from the present state to the next state, i.e.,

$$\pi_{ij} = \frac{\exp(\theta_{ij} + \boldsymbol{X\gamma}_{ij})}{\sum_{j=1}^{k} \exp(\theta_{ij} + \boldsymbol{X\gamma}_{ij})} \tag{7}$$

The transition model (3×3 probability matrix) contains three multinomial logistic regressions for each of the three states, and the probabilities are dependent on all six variables as in the FMR classification analysis. The initial state probabilities are also similarly determined by a multinomial logistic regression.

The parameter fitting process involves the backward-forward algorithm for likelihood calculation and the Baum-Welch method, an iterative procedure for optimization similar to the EM method. In addition to model coefficients, the optimal state sequence can also be estimated through the Viterbi algorithm (Zucchini et al. 2016). A more detailed description of the algorithms can be found in the statistical R-package *depmixS4* (2010: depmixS4: an R-package for hidden Markov models. **36**.). To prepare the temporal sequences for model fitting, it is intuitive to assume that each storm contributes to a single time sequence. However, since a TC behaves differently when over land, we filtered out these observations by dividing the storm's intensity evolution into multiple temporal sequences. That is to say, the time series may begin with either the genesis of the storm or the first observation after the storm makes landfall and moves back to the ocean again. The time series ends when the storm dies or hits land. However, the temporal sequences in between can be very short (sometimes less than 3 observations) when a storm hits land twice within a short period of time. These short sequences are considered as the "noise" of the training data and thus are not used. After removing sequences that are shorter than 12 observations (i.e., 3 days), 10072 observations, which consist of 357 time series, remain for model training.

Figure 1 the shows observed DV colored by estimated optimal states. The first state has a sample mean close to zero (-0.05) with a small sample standard deviation (0.50); the second state features a moderate positive mean (0.60) and variation (5.58), while the third state has a moderate mean (0.25) but the largest standard deviation (14.15) with DV taking large negative and positive values. The three regimes of intensification also appear clearly as three phases during the lifetime of individual storms. Also, as listed in Table 3a, the model standard deviations of the three states show similar patterns with the three subgroups in FMR (Table 2a). Similar explanations can be applied here so that the three underlying states represent storm's "static" (State 1), "moderate" (State 2), and "extreme" (State 3) intensification.

Other estimated parameters of the response component of the MeHiM are also shown in Table 3(a). Improved over the FMR results (Table 2a), the estimated coefficients are mostly with expected signs, except for V in State 1 and RH in States 1 and 3, though these coefficients are all relatively small in absolute value. Actually, all coefficients in State 1 are quite small compared with those in States 2 and 3. The expectation of DV (or the mean of the subgroup Gaussian distribution) in State 1 is almost 0 and is not sensitive to the change in the predictors. These parameters of the response model will be further examined together with the parameters of the transition model.

The estimated coefficients of the transition component of the MeHiM are shown in Table 3(b). Three multinomial logistic models are developed for each of three original states, and transitioning to State 3 is picked as the baseline (all coefficients set as 0). The signs and the relative magnitude of the regression coefficients jointly determine the change of the transition probabilities as a response to the change of the covariates. Consider the coefficients for storms that are currently in State 1 as an example. The linear combination of covariates describes the log odds of the storm staying in State 1 vs. transitioning to State 3 (first row) and that of the storm transitioning to State 2 vs. to State 3 (second row). For instance, a one-unit, or one standard deviation, increase in the variable $DV_p$ is associated with a decrease in the log odds of staying in State 1 vs. transitioning to State 3 in the amount of 0.103, which is equivalent to a decrease in the odds by approximately 10%. At the same time, an increase in the variable $DV_p$ is also associated with the increase in the odds of transitioning to State 2 vs. to State 3 by 5%. Thus, when all other covariates are fixed, an increase in $DV_p$ will cause storms in State 1 more likely to transition to State 2, and less likely to stay in State 1. Similarly, a one-unit increase in MPI will lead to a decrease in both the odds of staying in State 1 vs. transitioning to State 3 and the odds of transitioning to State 2 vs. to State 3. Thus storms are more likely to transition to State 3. In contrast, a one-unit increase in RH will lead to an increase in both of these odds and thus a decrease in probability to transition to State 3. The probabilities of both transitioning to State 2 and of staying in State 1 will grow, with the probability of transitioning to State 2 growing more rapidly than that of staying in State 1.

The effect of the change in each covariate on DV depends jointly on the transition probabilities and response distributions; this is particularly true for the MeHiM, where the two model components are coupled. Thus, one should combine Tables 3(a) and 3(b) to examine the environment's control on TC intensity change. We first consider the two storm-related variables $DV_p$ and V. As the coefficients in front of $DV_p$ in the response model are all positive, increases in $DV_p$ always lead to increases in DV, as expected. Specifically, for storms that are originally in States 2 and 3, an increase in $DV_p$ will result in an increase in the probability of transitioning to State 3, where the storm can significantly grow, as the response model coefficient in front of $DV_p$ is relatively large (~0.36). For storms that are originally from State 1 (storms that keep their intensity nearly as constant, with the response model coefficient in front of $DV_p$ as low as 0.01), an increase in $DV_p$ leads to a greater chance in transitioning to State 2, so that storms will start to develop quickly (with the response model coefficient in front of $DV_p$ as high as 0.6). Thus, the effect of $DV_p$ (storm inertia) is quite significant, although the response model coefficient in front of $DV_p$ appears to be low for State 1. The response model coefficient in front of V is negative for States 2 (-0.14) and 3 (-0.48), as expected, but it is marginally positive for State 1 (0.004). However, with larger V, storms that are in States 1 and 2 are more likely to transition to State 3. Only storms that are originally in State 3 are more likely to transition to State 1, where the nearly zero positive coefficient in front of V prevents a storm from growing, although it does not have an expected negative effect on the storm's intensification (possibly due to statistical compensation).

The model coefficients for the four environmental variables can be examined in a similar way. As the MPI increases, storms that are originally in States 1 or 2 are more likely to enter State 3 and hence will intensify more rapidly due to the larger, positive coefficient for State 3 (0.370 vs. 0.105 and 0.008) in the response model. However, for storms that are originally in State 3, a larger MPI may lead to the storm's transitioning to States 1 or 2, where the storm will achieve a smaller intensification. This mechanism may prevent storms from continuously intensifying at a large rate. In contrast, as SHR increases, storms that are originally in States 1 or 2 are more likely to enter State 3, and hence will deintensify more rapidly due to the larger, negative coefficient for State 3 (-0.619 vs. -0.044 and 0) in the response model. Storms that are originally in State 3 are more likely to transition to States 1 or 2, where the storm will achieve a relatively smaller deintensification. This mechanism may prevent storms from continuously deintensifying at a large rate. The response model coefficient in front to RH is positive (0.069) as expected for State 2, but it is moderately negative for States 1 (-0.007) and 3 (-0.023). However, as RH increases, storms in States 1 and 2 will be more likely to transition to State 2 (than to State 1). Only storms in State 3 will be more likely to be in State 3, where the unexpected, negative response in DV is likely due to statistical compensation to the large responses to other variables. The response model coefficients in front of OCN are all positive, with the coefficient for State 3 (0.252) larger than States 2 (0.039) and 1 (0.014). As OCN increases (ocean weakening effect decreases), storms at all states are more likely to enter State 3 and intensify more rapidly. The model seems to capture the expected large effect on storm intensification of ocean cooling, which will be discussed further in Section 4.

Finally, we consider the overall effects of the joint changes in the covariates. We estimate the transition probabilities for the three specific scenarios considered in the FMR classification analysis when all variables are at the 95$^{th}$ percentile of their samples that are unfavorable for larger DV, all variables are at their medians, and all variables are at the 95$^{th}$ percentile that are favorable for larger DV, as shown in Figure 2. We find that storms in State 1 are around 50% likely to stay in State 1 and around 50% likely to transition to State 2, regardless of the environment. In other words, storms in State 1 are likely to stay in State 1 (with small DV response to the environment) or move to State 2 (with moderate DV response to the environment). This mechanism makes small storms develop slowly. On the contrary, storms in States 2 and 3 respond to the environment in different directions. Moving from unfavorable conditions to favorable conditions, storms in State 2 are more and more likely to transition to State 1 (with probability changes from 5%-23%-29%), and less and less likely to stay in State 2 (94%-75%-69%). Storms that are originally in State 3 are significantly more and more likely to stay in State 3 (2%-50%-87%). Thus, although the probability of entering State 3 from States 1 (<1%) or 2 (<2%) is always low, once the storm enters State 3, it is likely to stay in State 3 and continuously achieve rapid intensification if the environment is favorable. This "lock-in" mechanism supports the development of intensity extremes, which is a significant advantage of the MeHiM over FMR. On the other hand, when the environment is unfavorable, the storms are not likely to enter State 3 or stay locked in State 3, and thus the model does not support much the development of rapid deintensification, similar to FMR.

2.4. Land Decay Model

TCs weaken quickly over land mainly due to the lack of fuel sources and the entrainment of dry air. Estimating the decay of storms after making landfall or when moving across small islands is also a key component of TC intensity models. Georgiou (1985) simulated the increase of central pressure as an inversely proportional function of distance from the landfall point, while Kaplan and DeMaria (1995) modeled the decay in the maximum wind speed as an exponential function of time after making landfall. DeMaria et al. (2006) proposed a similar model and applied it particularly for storms moving over narrow landmasses. The decay modes of Kaplan and DeMaria (1995) and DeMaria et al. (2006) have been applied to adjust operational forecasting, and verification on SHIPS has shown that the inclusion of such decay models can improve intensity forecasts in multiple basins (DeMaria et al. 2006).

Here, we develop a land model based on the same dataset applied for the over-ocean intensity models. The model structure is similar to that in Demaria and Kaplan (1995), except that we set the reduction factor to be 1 for simplicity, ignoring the assumed sudden intensity decrease when storms approach land. In this model, the storm's intensity $V$ exponentially decays as a function of time $t$, as shown in Eq. (6).

$$V(t) = V_b + (V_0 - V_b) \cdot e^{-\alpha t} \tag{8}$$

where $V_0$ is the storm's original intensity prior to landfall, which decays with a constant temporal rate $\alpha$, to a constant background wind, $V_b$, that a storm may maintain over land.

To estimate $\alpha$ and $V_b$ in this model, we collect all TC intensity sequences when the storm's center is over land. While most of these segments are quite short, with an average length of about 5 observations, we first exclude segments that are shorter than 2 observations (12 hours). Also, as the model does not apply for storms weaker than $V_b$, we assume the storms will not decay if the landfall intensity is smaller than $V_b$. Accordingly, we exclude segments with $V_0$ smaller than 20 kt with try and error. The model parameters $\alpha$ and $V_b$ are then fitted based on 229 segments, and the optimal decay rate and background wind are estimated to be 0.049 kt/6h and 18.82 kt, respectively, by nonlinear least squares regression. The fitted model predicts the intensity decay over land that compares well with the observation for storms with various initial intensities, as shown in Fig. 3.

3. Model Evaluation

To evaluate the developed models, we perform MC simulations of the intensity evolution over storms' lifetime for historical tracks and compare model simulations with observations. We examine the models' capability to reproduce the observed TC intensity metrics, including spatial distribution of intensity over the basin, probabilistic distributions of intensification rate, LMI, and landfall intensity, as well as temporal evolution of intensity for sample cases.

### 3.1. Monte Carlo (MC) simulation

For all simulations, the storm is initialized by the first two steps of observed intensity to obtain initial $V$ and $DV_p$. Then, for the OLS, the intensity change for each step is obtained by sampling from a Gaussian distribution (Eq. 2 and Table 1). For the FMR, the classification model is first applied to compute the mixture weights (Eq. 4 and Table 2b), based on which group membership is determined with the discrete acceptance-rejection method. The intensity change is then obtained by sampling the Gaussian distribution from the selected group (Eq. 3 and Table 2a).

Simulation with the MeHiM has one more step in the initialization. Given the environmental parameters (i.e., MPI, SHR, and RH) at the storm's occurrence, we use the initial state model with the acceptance-rejection method to determine the state for the first two steps. Then, we continue the simulation by repeatedly choosing the next state to visit according to the transition probabilities (Eq. 7 and Table 3b) with the discrete acceptance-rejection selection and updating storm intensity with the intensity change sampled from the state's corresponding response distribution (Eq. 6 and Table 2a).

During all simulations, the land model is applied when the storm's center hits land, and the over-the-ocean intensity models are resumed if the storm moves back to the ocean. Storms with $V_0$ smaller than $V_b$ are simulated to keep their original intensity over land, while storms with $V_0$ greater than $V_b$ are simulated to decay according to the filling model (Eq. 8). For the MeHiM, the storm's states over land are assumed to be the same as the last state when it is over ocean, before hitting the land. Simulations stop at the end of the historical record of the track or when the storm' intensity becomes lower than 10 kt. Each historical storm is simulated 100 times with each of the three intensity models, and the comparisons with observations are based on all realizations.

### 3.2. Spatial distributions

Fig. 4 compares observed and MeHiM-modeled TC intensity along historical tracks. The basin is divided as a 2°×2° grid and, as an example, the 90$^{th}$ percentile and 75$^{th}$ percentile of the intensity values in each grid from the observations are compared with those from simulations (median over the 100 realizations). The simulations capture relatively well the spatial variation of TC intensity, although they appear smoother and seem to underestimate the most extreme intensities over the lifetime of the storms (occurring over the Caribbean Sea and the Gulf of Mexico). Nevertheless, the MeHiM improves over the OLS and FMR in capturing the extremes (not shown here; see other comparisons below).

3.3. Probability distributions

In this subsection, we investigate the models' capability in capturing the probability distributions of various key intensity and intensity change metrics. First, a comparison of the histograms of observed (IBTrACS) and simulated 6-h intensity change rate (6-h DV) is shown in Fig. 6. Here observation and simulations over land are removed to focus on the comparison between the three over-the-ocean intensity models. All three models capture the general features of the observed distribution. However, simulations from the OLS have 6-h intensity changes fall within a narrow range of -26 ~ 22 kt. The FMR improves significantly over the OLS to a range of -46 ~ 44 kt. Both ranges are narrower than that of IBTrACS, which is -65~55 kt. The outlier at the right tail of IBTrACS (Fig. 5b) comes from an observation of 6-h DV of 55 kt during the lifecycle of Hurricane Wilma (2005), one of the most intense storms ever recorded in the Atlantic Basin. The outlier at the left tail comes from two observations of 6-h DV of about -65 k during the lifecycle of Hurricanes Iris (2001) and Dennis (2005). The MeHiM generates a large range of 6-h DV (-58 ~ 47kt), covering nearly the full range of observations except the outliers and improving over the OLS and FMR.

We also compare the histograms of 24-h intensity change (24-h DV) between observations and simulations (Fig. 6). The distribution from MeHiM simulations is more dispersed than that from the OLS and FMR and compares better with IBTrACS. In this case, the simulated range of 24h $DV$ is $-76$~63 kt for the OLS, $-114$~106 kt for FMR, and −100∼111 kt for MeHiM, compared to −120∼96 kt for IBTrACS. We find that although both the FMR and MeHiM are able to simulate extreme 24-h DV, simulations from FMR have fewer probability densities than IBTrACS in all bins that are larger than 22.5 kt. While simulations from the MeHiM have less density than IBTrACS in the interval of 22.5∼37.5 kt, they have larger densities in the intervals greater than 37.5 kt. Particularly, although the MeHiM does not capture the outlier extreme 6-h intensification (Fig. 5b), it produces larger, more extreme 24-h intensifications and higher probability densities for the bins of 24h DV greater than 30 kts (i.e., RI) than IBTrACS (Fig. 6b). On the other hand, the MeHiM underestimates both the rates of most extreme 6-h (Fig. 5b) and 24-h (Fig. 6b) deintensifications. The large 24-h intensifications are likely generated by the "lock-in" mechanism in the MeHiM, which supports continuous large intensification once the storm enters the extreme state and the environment is favorable. The MeHiM lacks such a mechanism to support continuous large deintensification, as discussed earlier.

Next, we compare the probability distribution of lifetime maximum intensity (LMI) in the simulations and observations (Fig. 7). Among the three models, the MeHiM compares with IBTrACS the best. It captures the shape and peak of the LMI distribution quite well, although it is smoother and misses the "fluctuations and bumps" in the tail, and as a result it underestimates the extreme LMIs to some extent. The OLS and FMR estimated LMI distributions have significantly thinner tails. Also, FMR-estimated distribution has a substantially higher peak, and OLS-estimated simulation has a significant right shift in the peak, compared to IBTrACS and the MeHiM. Thus, the OLS and FMR are likely to underestimate the LMI for extreme storms and overestimate the LMI for relatively weak storms.

To further investigate the matter, we examine the LMI distribution for non-RI and RI storms separately. The percentages of simulated RI storms by the OLS, FMR, and MeHiM are 31.0%, 20.9%, and 25.6%, respectively, while in IBTrACS this fraction is 24.9%. Thus, the MeHiM improves over the OLS and FMR in simulating the right fraction of RI storms. Furthermore, as shown in Fig. 8, the two LMI distributions of the OLS simulations are quite close, so that the peak in the LMI distribution of non-RI storms is shifted to the right while the peak in the LMI distribution of RI storms is shifted to the left, compared to IBTrACS. Therefore, the OLS tends to overestimate the LMI of non-RI storms and to underestimate the LMI of RI-storms (although it overestimates the faction of RI storms). FMR better captures the location of the peak of the distribution for non-RI storms, but it overestimates the peak. Also, similar to the OLS, the LMI distribution for RI-storms in FMR is left-shifted and featured with a much thinner tail compared to IBTrACS. Thus, FMR tends to underestimate the LMI for RI storms (and also underestimate the likelihood of RI). Significantly improving over the OLS and FMR, the MeHiM captures the LMI distribution of non-RI storms very well. The MeHiM also captures the shape of the LMI distribution for RI storms much better, although the estimated distribution is still slightly left-shifted and thus the most extreme LMIs for RI storms are still underestimated.

Comparison between Figs. 7 and 8 indicates that the tail of the LMI distribution is largely contributed by RI storms, as noted by Lee et al. (2016a), and thus the statistical intensity models underestimate the tail of the LMI distribution mainly because they underestimate the LMI of extreme RI storms. Given that 80% of major TCs (LMI > 96kt) underwent RI in history, it is important to capture the RI distribution in order to capture the LMI distribution. Being able to capture the distribution of RI extremes, however, does not imply being able to capture the tail of the LMI distribution. Improving over the OLS and FMR, the MeHiM captures and even slightly overestimates the tail of the 24-h DV distribution representing RI (Fig. 6b), but it still underestimates the tail of the LMI distribution (Figs. 7 and 8). This underestimation of extreme LMIs also explains the underestimation of spatial extremes in Fig. 4.

Then, we compare the simulated and observed probability distribution of the landfall intensity for the North Atlantic coastline (Fig. 9). Both the OLS and MeHiM capture the landfall probability distribution well, with the MeHiM slightly better in capturing the tail. FMR underestimates the tail more than the MeHiM and OLS and significantly overestimates the peak. Also, while the $15^{th}$-$85^{th}$ percentile bond of the OLS and MeHiM estimates can generally cover the distribution from IBTrACS, the percentile bond of FMR estimates is quite narrow, and it often misses IBTrACS. It is noted that for the Northeastern US coastline, the distribution from IBTrACS is quite rough and the percentile bond of the model estimates is relatively large, due to data limitations for both the observations and simulations for this region.

3.4. Sample cases

Apart from capturing various statistical distributions, we compare the performances of the OLS, FMR, and MeHiM in simulating the intensity evolution of individual storms. Figure 10(a-c) show as an example the intensity evolution of a normal TC, Hurricane Alex in 2010, simulated by the OLS, FMR, and MeHiM, compared to the IBTrACS observation. All three models can capture the main features of the storm's development at the early stage and its decay to death in the end. The ensemble mean (over the 100 realizations) from each model compares well with the observation. The MeHiM performs slightly better as the LMI of the ensemble mean shifts toward the observed LMI, compared to the OLS and FMR.

The intensity simulations for an extreme RI storm, Hurricane Wilma (2005), are shown in Fig. 10 (d-f). Hurricane Wilma experienced the most intense rapid intensification (96kt in 24 hours) over the period of 1979-2014 in the North Atlantic Basin and reached an LMI of 160kt. The LMI of the ensemble mean of the models is significantly lower than the observation, with the MeHiM performing slightly better. This is somewhat expected given the models' underestimation of the tail of the LMI distribution (see Figs. 7 and 8). However, even for such an extreme case, we find that the ensemble of the MeHiM is able to cover the observed LMI – the maximum (over the 100 realizations) of the simulated LMI is 175 kt from the MeHiM, while it is 135 kt from OLS and 147 kt from FMR. In addition, about 40% of the realizations from the MeHiM can reach a LMI over 100 kt, which is much more than that from OLS and FMR (~15%). Indeed, an upward shift over the early development stage in the MeHiM realizations (indicated by the shadings) indicates a larger simulated intensification rate in the MeHiM compared to the OLS and FMR. Hurricane Wilma weakened greatly in the later stage of its lifecycle, when its intensity decreased by 35 kts within 24 hours. The extreme cases of the MeHiM realizations (shading) are able to represent this rapid deintensification and large decay.

In Fig. 9(g-i), an unnamed Tropical Depression in 2003 is shown as an example of weak storms with LMI less than 45 kt. In this case, both FMR and the MeHiM are capable of simulating the intensity evolution, with the ensemble mean matching the observation. The OLS, on the other hand, significantly overestimates storm's intensity, with mean as well as the $20^{th}$ -$80^{th}$ percentile envelope shift upward away from the observation. This overestimation by the OLS is typical for weak storms, as discussed before (see Figs. 7 and 8).

From the above results, we find that both FMR and the MeHiM are able to simulate relatively weak storms, while the OLS tends to overestimate weak storms. The MeHiM has significant advantage over both FMR and the OLS when simulating RI storms or storms with large intensification or deintensification. It is also noted that the models tend to overestimate the storm's intensity after making landfall (e.g., Figs. 10a-c), which indicates that the simplified land model may not be sufficient to simulate a storm's rapid decay over land.

4. Discussion

    4.1. Effect of the ocean feedback variable

The three (over-ocean) intensity models are developed with four environmental covariates including three atmospheric and one ocean feedback variable. As the ocean feedback variable (OCN) is not used in previous statistical models, here we test its significance and examine how OCN represents TC-ocean interaction. We use two storms as examples: Tropical Storm Larry (2003) and Hurricane Felix (1995). The upper two panels in Fig. 11 show simulated intensity evolution for these two storms from the original MeHiM (as discussed above), and the lower two panels show simulations from the MeHiM without OCN (rebuilt with OCN removed). The simulations of MeHiM shift toward the observation greatly compared to the MeHiM without OCN for Tropical Storm Larry, which implies OCN's significant role in preventing simulated storms from growing too fast. The OCN generally captures the negative effect of the TC-ocean interaction on the storm's intensification, i.e., the stronger the storm, the stronger the ocean's cooling effect. It is more interesting to note, however, that including OCN may also better simulate extreme intensification. For Hurricane Felix, an RI storm in 1995, simulations from the MeHiM with OCN are better in terms of capturing the rapid RI phase, compared to the simulations from the MeHiM without OCN. In this case, the relatively weak effect of the ocean feedback during RI is captured by OCN. Thus, adding OCN does not always make storms weaker, and it may also help simulate storm's rapid intensification in a statistical model.

### 4.2. Effect of extreme state identification

The MeHiM performs well for simulating normal and relatively weak storms but may underestimate extremes (as shown in Figs. 7, 8, and 10). This underestimation of a storm's LMI is often due to the underestimation of rapid or extreme intensification, which in turn is likely due to the limitation in capturing the underlying extreme state of intensity change (i.e., underestimating the probability of entering the extreme state, as discussed in Lin et al. 2017 for the FMR analysis). Here we investigate this limitation in the MeHiM by examining four extreme RI storms with maximum intensification rate ranging from 35 kt to 85 kt within 24 hours. For each of these storms, we run the MeHiM freely until the storm experiences rapid intensification, when the occurrence of RI is indicated by IBTrACS. At the time of RI, the underlying state of intensity change is manually fixed to the extreme state (regardless of the transition probability in the model) during the next four steps (24 hours, consistent with the definition of RI). Thereafter, the model is resumed to run freely. The results are shown, in comparison with the original MeHiM runs, in Fig. 12, where the black arrow indicates the occurrence of RI in each case. We find that the simulations improve significantly with the correction of the underlying states at the occurrence of RI. This is especially the case for Hurricane Andrew (1992). The MeHiM performs well in simulating the first part of the storm's intensity evolution, but it fails to capture the RI and thus greatly underestimates the peak intensity after RI. With state correction, the simulations are shifted greatly towards observation and thus well capture the observed LMI. The rapid decay after landfall is again not well captured for these extreme cases due to the limitations of the land model.

We also evaluate the MeHiM's performance and limitation in simulating three recent destructive hurricanes in the 2017 Atlantic hurricane season, i.e, Hurricanes Harvey (formed on Aug $17^{th}$), Irma (formed on Aug $30^{th}$), and Maria (formed on Sep $16^{th}$), which are all RI storms. As a comparison, results from OLS are shown in Fig. 13 (a-c). Fig. 13 (d-f) shows simulated results from the original MeHiM. The MeHiM moderately improves over OLS but still significantly underestimates the LMI of these extreme RI storms. However, when the prior knowledge of RI is given, the simulations from the MeHiM with state correction improve greatly. According to the observation, Harvey began to undergo RI in the morning of August 24, the RI of Maria occurred on Sep 17$^{th}$, and Irma is a special case since it experienced two periods of RI during its lifecycle, one at the very beginning, on Aug 31$^{st}$, and the other on Sep 4$^{th}$. The MeHiM simulations with state correction for these RI periods, as shown in Fig. 13(g-i), compare much better with the observations, especially for Hurricane Maria and the first RI of Hurricane Irma.

This analysis confirms that although the MeHiM improves over FMR in capturing the extreme state, it may still underestimate the likelihood of extreme states. Future research may focus on improving the state transition probability modeling of the MeHiM. This analysis also indicates that the MeHiM may have a great potential for real-time operational forecasting, since, unlike a homogenous model such as the OLS, the state of intensity change in the MeHiM can be manipulated through incorporating/assimilating real-time information on the RI state, e.g., the probabilistic guidance RII (Kaplan et al. 2010).

5. Summary

In this study, we have developed a Markov environment-dependent hurricane intensity model, or the MeHiM, based on the North Atlantic TC records from IBTrACS WMO archive and environmental parameters derived from ERA-Interim reanalysis data over the period of 1979-2014. The model considers three hidden discrete states, and each state is associated with a probability distribution of intensity change. The movement of intensity change from one state to another is considered a Markov chain described by a transition probability matrix. Both the intensity change and state transition components of the model are dependent on environmental variables (including potential intensity, vertical wind shear, high-level relative humidity, and ocean feedback) and storm variables (including current intensity and previous intensity change). All the variables, including the ocean feedback variable that is first applied in statistical modeling, are found to be significant. Similar to the FMR model discussed in Lin et al. (2017), the MeHiM's three states of intensity change turn out to represent the storm's slow, moderate, and rapid intensity change, respectively. The MeHiM improves over FMR as it accounts for the temporal correlation of intensity change states and integrates the intensity change and state identification components in a Markov framework. As a result, the MeHiM significantly improves over FMR in better capturing the extreme state and simulating rapid and continuous intensification.

We evaluate the MeHiM by comparing simulated historical storms with observations, where the influence of land is included through a simplified land decay model. We find that the MeHiM shows a great improvement over previous statistical models (such as the OLS and FMR, built as baseline models in this study) in simulating TC intensity climatology. The probability distributions of TC intensification rates (6-h DV and 24-h DV) and LMI simulated by the MeHiM are closer to the observations. With the capability in simulating Category 4 and 5 storms, the MeHiM can better capture the tail of the LMI distribution, although it still slightly underestimates the tail. The MeHiM is limited in simulating the most extreme LMIs, even though it can well capture the tail of the 24-h DV distribution representing RI, indicating that a mechanism of continuous intensification beyond the temporal window of RI (i.e., 24 hours) exists but is not well captured by the model. The MeHiM also improves over previous models in reproducing the observed distribution of landfall intensity for regions with relatively adequate data. The model can also be applied to provide advisory to areas with limited historical data.

The MeHiM improves over the OLS and FMR in probabilistically forecasting the temporal evolution of intensity for individual storms, although the MeHiM still may not capture a storm's RI, especially for storms that strengthen extremely in a short time period. Nevertheless, we find that, when combined with an extreme state correction for RI, the forecasting can be greatly improved for most historical RI storms. We demonstrate this feature of the MeHiM for various extreme cases, including the three most recent ones, Hurricanes Harvey, Irma, and Maria in the 2017 Atlantic hurricane season. This analysis indicates the potential to significantly improve the MeHiM through improving its state transition probability model to better capture the extreme state. This analysis also indicates a potential application of the MeHiM combined with a probabilistic RI guidance in real-time operational forecasting.

Future studies may also include a comparison between the MeHiM and dynamic models, such as the Coupled Hurricane Intensity Prediction System (CHIPS; Emanuel et al. 2004) and its simplified algorithm (Emanuel 2017). The ultimate goal of our study is to integrate the MeHiM with a genesis model and a track model into a complete hurricane climatology modeling system, similar to Emanuel et al. (2008) and Lee et al. (2018).

Acknowledgments

This study is supported by Grants 1652448 from the National Science Foundation and NA14OAR4320106 from the National Oceanic and Atmospheric Administration, U.S. Department of Commerce. The statements, findings, conclusions, and recommendations herein are those of the authors and do not necessarily reflect the views of the National Science Foundation, the National Oceanic and Atmospheric Administration, or the U.S. Department of Commerce.

References

Balaguru, K., G. R. Foltz, L. R. Leung, E. D. Asaro, K. A. Emanuel, H. Liu, and S. E. Zedler, 2015: Dynamic Potential Intensity: An improved representation of the ocean's impact on tropical cyclones. *Geophysical Research Letters*, **42**, 6739–6746, doi:10.1002/2015GL064822.

Balmaseda, M. A., K. Mogensen, and A. T. Weaver, 2013: Evaluation of the ECMWF ocean reanalysis system ORAS4. *Quarterly Journal of the Royal Meteorological Society*, **139**, 1132–1161, doi:10.1002/qj.2063.

Batts, M. E., E. Simiu, and L. R. Russell, 1980: Hurricane Wind Speeds in the United States. *Journal of the Structural Division*, **106**, 2001–2016.

Bister, M., and K. A. Emanuel, 1998: Dissipative heating and hurricane intensity. *Meteorol Atmos Phys*, **65**, 233–240, doi:10.1007/BF01030791.

Bister, M., and K. A. Emanuel, 2002: Low frequency variability of tropical cyclone potential intensity 1. Interannual to interdecadal variability. *Journal of Geophysical Research: Atmospheres (1984–2012)*, **107**, ACL26–1–ACL26–15, doi:10.1029/2001JD000776.

Camargo, S. J., and A. A. Wing, 2016: Tropical cyclones in climate models. *Wiley Interdisciplinary Reviews: Climate Change*, **7**, 211–237, doi:10.1002/wcc.373. http://onlinelibrary.wiley.com/doi/10.1002/wcc.373/full.

Cappé, O., E. Moulines, and T. Ryden, 2006: *Inference in Hidden Markov Models*. Springer Science & Business Media, 1 pp.

de Boyer Montégut, C., G. Madec, A. S. Fischer, A. Lazar, and D. Iudicone, 2004: Mixed layer depth over the global ocean: An examination of profile data and a profile-based climatology. *Journal of Geophysical Research: Atmospheres (1984–2012)*, **109**, 1521, doi:10.1029/2004JC002378.

Dee, D. P., and Coauthors, 2011: The ERA-Interim reanalysis: configuration and performance of the data assimilation system. *Quarterly Journal of the Royal Meteorological Society*, **137**, 553–597, doi:10.1002/qj.828.

DeMaria, M., and J. Kaplan, 1999: An updated Statistical Hurricane Intensity Prediction Scheme (SHIPS) for the Atlantic and eastern North Pacific basins. *Weather and Forecasting*, **14**, 326–337.

DeMaria, M., and J. Kaplan, 1994: A Statistical Hurricane Intensity Prediction Scheme (SHIPS) for the Atlantic Basin. *http://dx.doi.org/10.1175/1520-0434(1994)009<0209:ASHIPS>2.0.CO;2*, **9**, 209–220, doi:10.1175/1520-0434(1994)009<0209:ASHIPS>2.0.CO;2.

DeMaria, M., J. A. Knaff, and J. Kaplan, 2006: On the Decay of Tropical Cyclone Winds Crossing Narrow Landmasses. *http://dx.doi.org/10.1175/JAM2351.1*, **45**, 491–499, doi:10.1175/JAM2351.1.

Elsberry, R. L., T. D. B. Lambert, and M. A. Boothe, 2007: Accuracy of Atlantic and Eastern North Pacific Tropical Cyclone Intensity Forecast Guidance. *http://dx.doi.org/10.1175/WAF1015.1*, **22**, 747–762, doi:10.1175/WAF1015.1.

Emanuel, K., 2017: A fast intensity simulator for tropical cyclone risk analysis. *Nat Hazards*, **50**, 1–18, doi:10.1007/s11069-017-2890-7.

Emanuel, K. A., 1995: Sensitivity of Tropical Cyclones to Surface Exchange Coefficients and a Revised Steady-State Model incorporating Eye Dynamics. *J. Atmos. Sci*, **52**, 3969–3976, doi:10.1175/1520-0469(1995)052<3969:SOTCTS>2.0.CO;2.

Emanuel, K., C. DesAutels, C. Holloway, and R. Korty, 2004: Environmental control of tropical cyclone intensity. *J. Atmos. Sci*, **61**, 843–858.

Emanuel, K., R. Sundararajan, J. Williams, and J. Williams, 2008: Hurricanes and Global Warming: Results from Downscaling IPCC AR4 Simulations. *Bull. Amer. Meteor. Soc*, **89**, 347–368, doi:10.1175/BAMS-89-3-347.

Georgiou, P. N., 1985: *Design Wind Speeds in Tropical Cyclone-prone Regions*. 1 pp.

Kaplan, J., and M. DeMaria, 1995: A Simple Empirical-Model for Predicting the Decay of Tropical Cyclone Winds After Landfall. *Journal of Applied Meteorology*, **34**, 2499–2512.

Kaplan, J., and M. DeMaria, 2003: Large-scale characteristics of rapidly intensifying tropical cyclones in the North Atlantic basin. *Weather and Forecasting*, **18**, 1093–1108.

Kaplan, J., M. Demaria, and J. A. Knaff, 2010: A Revised Tropical Cyclone Rapid Intensification Index for the Atlantic and Eastern North Pacific Basins. *http://dx.doi.org/10.1175/2009WAF2222280.1*, **25**, 220–241, doi:10.1175/2009WAF2222280.1.

Knapp, K. R., M. C. Kruk, D. H. Levinson, H. J. Diamond, and C. J. Neumann, 2010: The International Best Track Archive for Climate Stewardship (IBTrACS). *http://dx.doi.org/10.1175/2009BAMS2755.1*, doi:10.1175/2009BAMS2755.1.

Knutson, T. R., and Coauthors, 2010: Tropical cyclones and climate change. *Nature Geoscience*, **3**, 157–163, doi:10.1038/ngeo779.

Lee, C.-Y., M. K. Tippett, A. H. Sobel, and S. J. Camargo, 2016a: Rapid intensification and the bimodal distribution of tropical cyclone intensity. *Nature Communications*, **7**, 10625, doi:10.1038/ncomms10625.39

Lee, C.-Y., M. K. Tippett, A. H. Sobel, and S. J. Camargo, 2016b: Autoregressive Modeling for Tropical Cyclone Intensity Climatology. *http://dx.doi.org/10.1175/JCLI-D-15-0909.1*, **29**, 7815–7830, doi:10.1175/JCLI-D-15-0909.1.

Lee, C.-Y., M. K. Tippett, S. J. Camargo, and A. H. Sobel, 2015: Probabilistic Multiple Linear Regression Modeling for Tropical Cyclone Intensity. *http://dx.doi.org/10.1175/MWR-D-14-00171.1*, **143**, 933–954, doi:10.1175/MWR-D-14-00171.1.

Lin, N., and D. Chavas, 2012: On hurricane parametric wind and applications in storm surge modeling. *Journal of Geophysical Research: Atmospheres (1984–2012)*, **117**, D09120, doi:10.1029/2011JD017126.

Lin, N., R. Jing, Y. Wang, E. Yonekura, J. Fan, and L. Xue, 2017: A Statistical Investigation of the Dependence of Tropical Cyclone Intensity Change on the Surrounding Environment. *Mon. Wea. Rev*, **145**, 2813–2831, doi:10.1175/MWR-D-16-0368.1.

Lloyd, I. D., and G. A. Vecchi, 2011: Observational Evidence for Oceanic Controls on Hurricane Intensity. *http://dx.doi.org/10.1175/2010JCLI3763.1*, **24**, 1138–1153, doi:10.1175/2010JCLI3763.1.

M. Demaria, and J. Kaplan, 1995: A Simple Empirical Model for Predicting the Decay of Tropical Cyclone Winds after Landfall. *http://dx.doi.org/10.1175/1520-0450(1995)034<2499:ASEMFP>2.0.CO;2*, **34**, 2499–2512, doi:10.1175/1520-

0450(1995)034<2499:ASEMFP>2.0.CO;2.

M. Mainelli, L. K. Shay, J. Kaplan, M. Demaria, M. Mainelli, and J. A. Knaff, 2005: Further Improvements to the Statistical Hurricane Intensity Prediction Scheme (SHIPS). *http://dx.doi.org/10.1175/WAF862.1*, **20**, 531–543, doi:10.1175/WAF862.1.

McLachlan, G., and T. Krishnan, 2007: *The EM Algorithm and Extensions*. John Wiley & Sons, Hoboken, NJ, USA, 1 pp.

Rabiner, L. R., 1989: A tutorial on hidden Markov models and selected applications in speech recognition. *Proc. IEEE*, **77**, 257–286, doi:10.1109/5.18626.

Rappaport, E. N., and Coauthors, 2009: Advances and Challenges at the National Hurricane Center. *http://dx.doi.org/10.1175/2008WAF2222128.1*, **24**, 395–419, doi:10.1175/2008WAF2222128.1.

Schade, L. R., and K. A. Emanuel, 1999: The Ocean's Effect on the Intensity of Tropical Cyclones: Results from a Simple Coupled Atmosphere–Ocean Model. *http://dx.doi.org/10.1175/1520-0469(1999)056<0642:TOSEOT>2.0.CO;2*, **56**, 642–651, doi:10.1175/1520-0469(1999)056<0642:TOSEOT>2.0.CO;2.

Tang, B., and K. Emanuel, 2012: A Ventilation Index for Tropical Cyclones. *http://dx.doi.org/10.1175/BAMS-D-11-00165.1*, **93**, 1901–1912, doi:10.1175/BAMS-D-11-00165.1.

Toepfer, F., R. Gall, F. Marks, E. R. H. Doc, 2010, *Hurricane Forecast Improvement*

*Program five year strategic plan*.

Vincent, E. M., M. Lengaigne, J. Vialard, G. Madec, N. C. Jourdain, and S. Masson, 2012: Assessing the oceanic control on the amplitude of sea surface cooling induced by tropical cyclones. *Journal of Geophysical Research: Atmospheres (1984–2012)*, **117**, n/a–n/a, doi:10.1029/2011JC007705.

Walsh, K. J. E., and Coauthors, 2016: Tropical cyclones and climate change. *Wiley Interdisciplinary Reviews: Climate Change*, **7**, 65–89, doi:10.1002/wcc.371.

Whitney, L. D., and J. S. Hobgood, 1997: The Relationship between Sea Surface Temperatures and Maximum Intensities of Tropical Cyclones in the Eastern North Pacific Ocean. *J. Climate*, **10**, 2921–2930, doi:10.1175/1520-0442(1997)010<2921:TRBSST>2.0.CO;2.

Willoughby, H. E., J. A. Clos, M. G. Shoreibah, J. A. Clos, and M. G. Shoreibah, 1982: Concentric Eye Walls, Secondary Wind Maxima, and The Evolution of the Hurricane vortex. *http://dx.doi.org/10.1175/1520-0469(1982)039<0395:CEWSWM>2.0.CO;2*, **39**, 395–411, doi:10.1175/1520-0469(1982)039<0395:CEWSWM>2.0.CO;2.

Zucchini, W., I. L. MacDonald, and R. Langrock, 2016: *Hidden Markov models for time series: an introduction using R*.

Table List:

Table 1. Coefficients of OLS model.

| Intercept | $DV_p$ | V | MPI | SHR | RH | OCN | $\sigma$ |
|---|---|---|---|---|---|---|---|
| -0.000 | 0.452 | -0.142 | 0.083 | -0.054 | 0.042 | 0.032 | 0.860 |

Table 2(a). Coefficients of FMR response model.

| | Weights | Intercept | $DV_p$ | V | MPI | SHR | RH | OCN | $\sigma_r$ |
|---|---|---|---|---|---|---|---|---|---|
| Group 1 | 0.296 | -0.080 | 0.008 | 0.003 | 0.008 | -0.004 | -0.008 | 0.012 | 0.098 |
| Group 2 | 0.635 | 0.100 | 0.667 | -0.134 | 0.116 | -0.063 | 0.063 | 0.055 | 0.715 |
| Group 3 | 0.069 | 0.228 | 0.215 | -0.861 | -0.133 | -0.646 | 0.153 | 0.234 | 2.029 |

Table 2(b). Coefficients of FMR classification model.

| | Intercept | $DV_p$ | V | MPI | SHR | RH | OCN |
|---|---|---|---|---|---|---|---|
| Group 1 | 3.053 | -0.359 | -1.091 | 0.020 | -0.260 | -0.274 | -0.217 |
| Group 2 | -0.292 | -0.292 | -0.777 | -0.148 | -0.212 | -0.101 | -0.140 |
| Group 3 | 0 | 0 | 0 | 0 | 0 | 0 | 0 |

Table 3(a). Coefficients of MeHiM response model.

| | Intercept | $DV_p$ | V | MPI | SHR | RH | OCN | $\sigma$ |
|---|---|---|---|---|---|---|---|---|
| State 1 | -0.077 | 0.010 | 0.004 | 0.008 | 0.000 | -0.007 | 0.014 | 0.097 |
| State 2 | 0.072 | 0.593 | -0.143 | 0.105 | -0.044 | 0.069 | 0.039 | 0.722 |
| State 3 | 0.269 | 0.357 | -0.483 | 0.370 | -0.619 | -0.023 | 0.252 | 1.979 |

Table 3(b). Coefficients of MeHiM transition model.

|  |  | Intercept | $DV_p$ | V | MPI | SHR | RH | OCN |
|---|---|---|---|---|---|---|---|---|
| From State 1 | To State 1 | 4.663 | -0.103 | -1.233 | -2.713 | -0.806 | 0.022 | -0.670 |
|  | To State 2 | 4.553 | 0.047 | -1.279 | -2.939 | -0.830 | 0.183 | -0.807 |
|  | To State 3 | 0 | 0 | 0 | 0 | 0 | 0 | 0 |
| From State 2 | To State 1 | 1.991 | -0.642 | -2.180 | -1.053 | -0.210 | 0.058 | -0.298 |
|  | To State 2 | 3.561 | -0.661 | -1.116 | -1.245 | -0.242 | 0.339 | -0.007 |
|  | To State 3 | 0 | 0 | 0 | 0 | 0 | 0 | 0 |
| From State 3 | To State 1 | -1.013 | -0.194 | 0.043 | 0.705 | 0.194 | -0.904 | -0.870 |
|  | To State 2 | 0.267 | -0.020 | -0.503 | 0.506 | 0.914 | -0.580 | -1.513 |
|  | To State 3 | 0 | 0 | 0 | 0 | 0 | 0 | 0 |

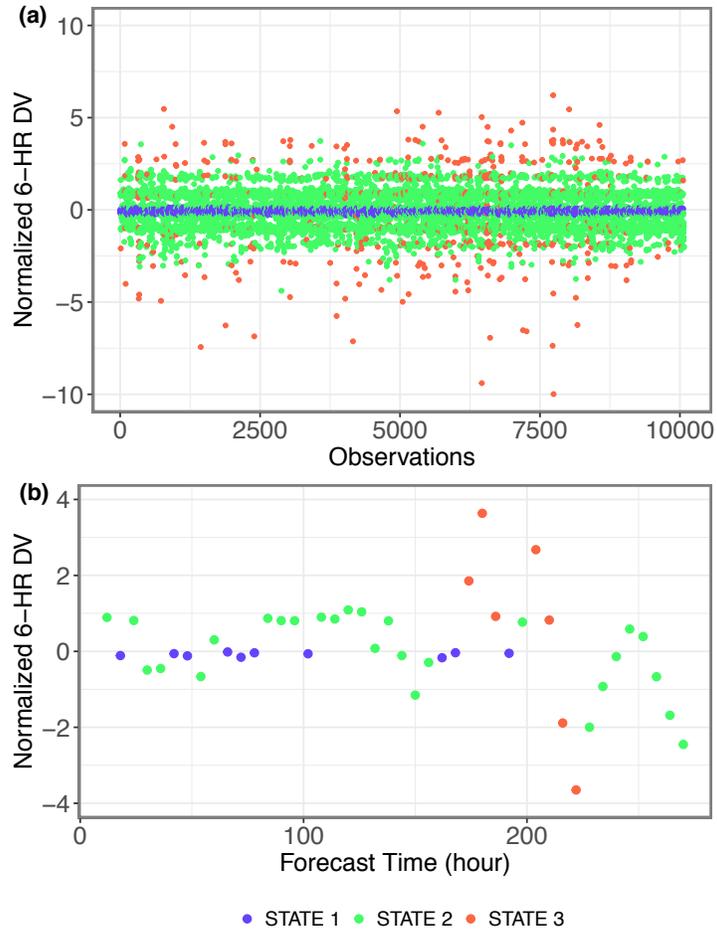

Fig 1. MeHiM optical states for (a) all data and (b) a sample storm. All observations of DV are plotted with colors denoting the optimal states given by MeHiM fitting.

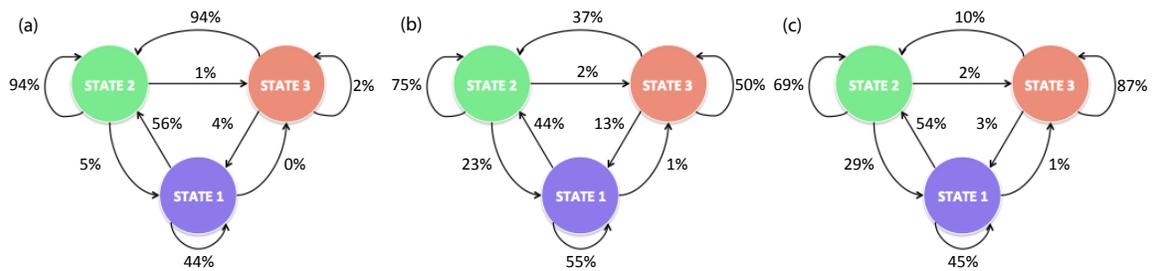

Fig. 2. Examples of MeHiM transition model when all covariates are (a) at 95$^{th}$ percentile of their samples that are unfavorable for larger DV, (b) at their medians, and (c) at 95$^{th}$ percentile of their samples that are favorable for larger DV. The transition probabilities are shown by the numbers. Variables in (a) are $DV_p$= 0.01 kt/6h, V = 34.4 kt, MPI = 141.0 kt, SHR = 14.2 m/s, RH = 50.0 %, OCN = 0.933. Variables in (b) are $DV_p$= 9.61 kt/6h, V = 15.3 kt, MPI = 161.6 kt, SHR = 8.3 m/s, RH = 69.5%, OCN = 0.999. Variables in (c) are $DV_p$= 9.14 kt/6h, V = 94.4 kt, MPI = 24.48 kt, SHR = 27.63 m/s, RH = 37.73%, and OCN = 0.398.

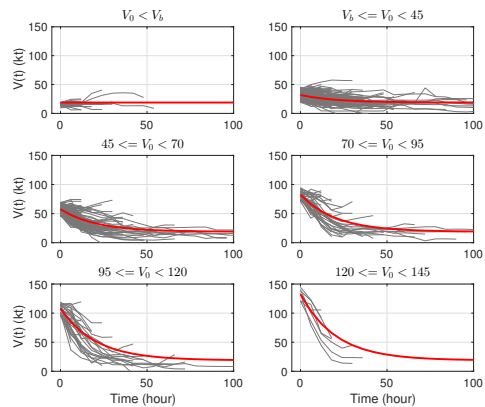

Fig 3. Comparison of land decay model (red curves) and observed intensity evolutions (grey curves) for different initial intensity ranges. The modeled decay is calculated with an initial intensity equal to the midpoint value of the intensity range.

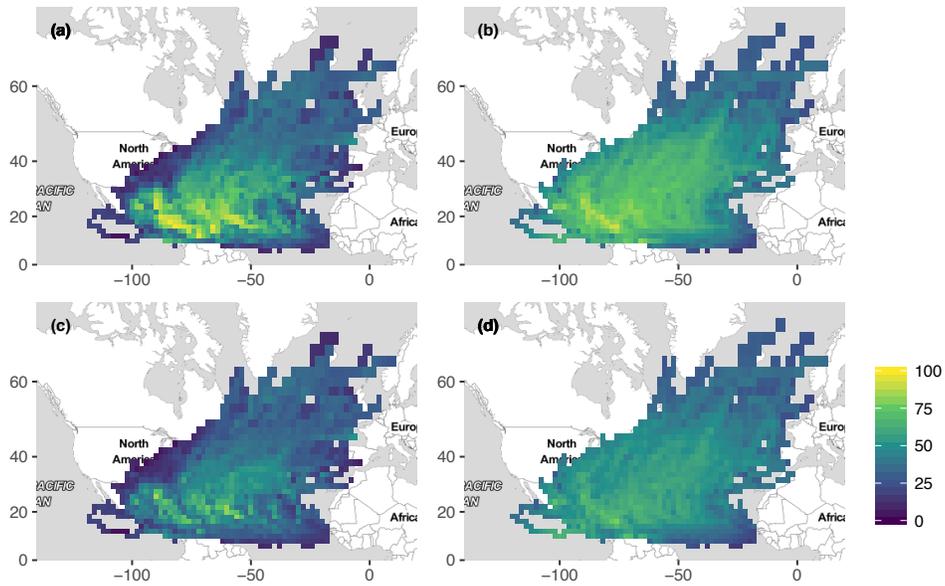

Fig. 4. Comparison of observed and MeHiM-modeled TC intensity (kt; color) in each 2°×2° grid. The intensity metrics plotted are (a) 90$^{th}$ percentile of observations, (b) median (over 100 realizations) of 90$^{th}$ percentile of MeHiM simulations, (c) 75$^{th}$ percentile of observations, and (d) median of 70$^{th}$ percentile of MeHiM simulations.

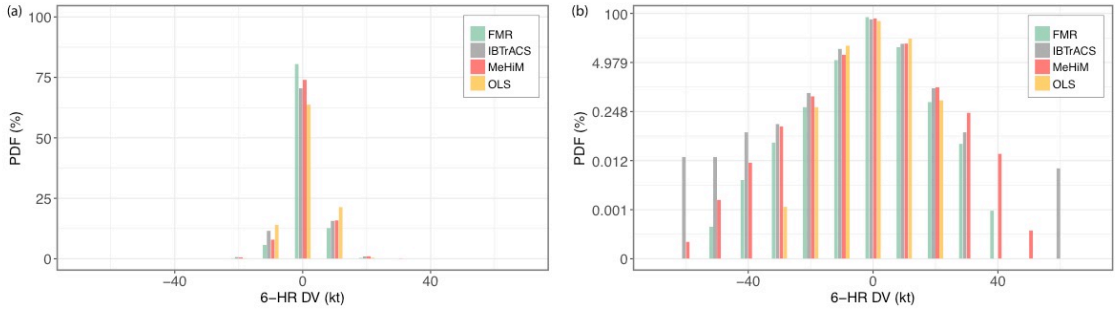

Fig. 5. Comparison of histograms of observed (grey) and modeled (OLS in yellow, FMR in green, and MeHiM in red) 6-h intensity change in (a) linear scale and (b) logarithmic scale. Over-land points are removed. Data are binned in 10-kt interval. Simulation results include 100 realizations.

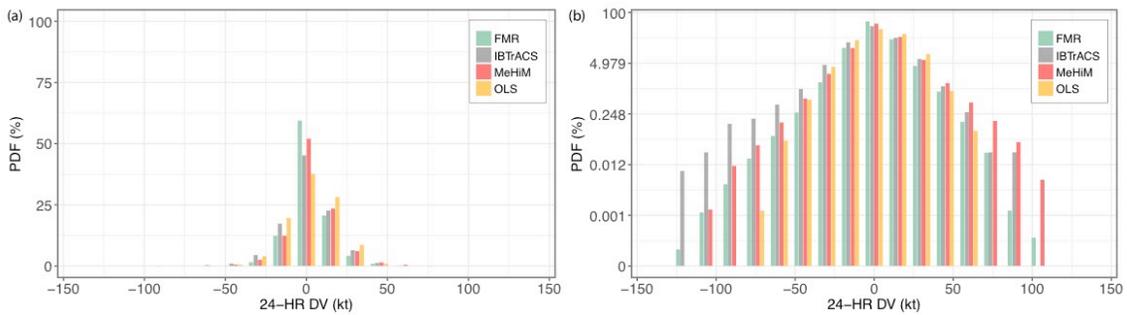

Fig. 6. Same as Fig. 5, but for 24-h intensity change. Data are binned in 15-kt interval.

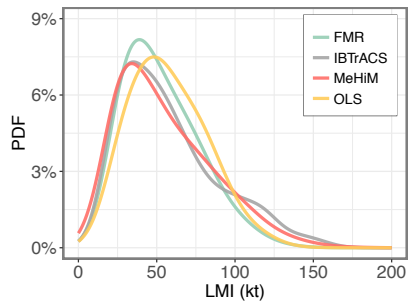

Fig. 7. Comparison of PDF of lifetime maximum intensity between observations (grey) and model simulations (OLS in yellow, FMR in green, and MeHiM in red). Over-land points are removed. Simulation results include 100 realizations.

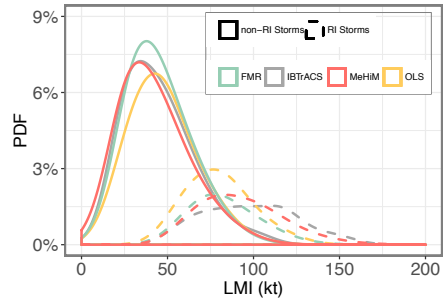

Fig. 8. Same as Fig. 7, but for RI storms (dashed curves) and non-RI storms (solid curves).

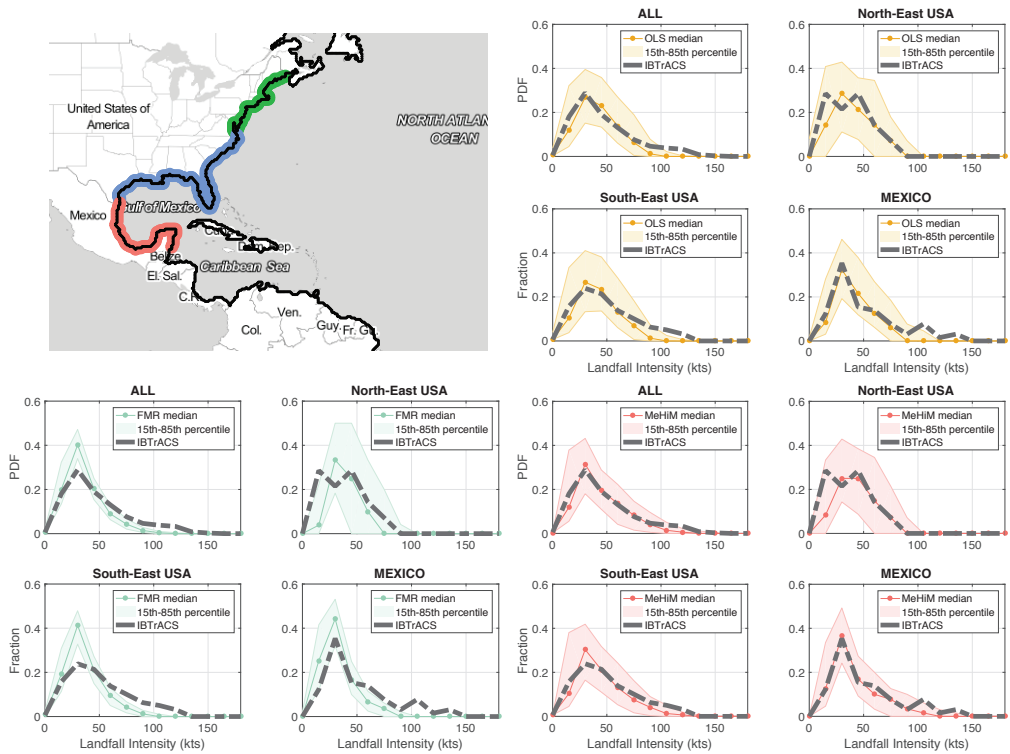

Fig. 9. Comparison of observed (thick dashed curves) and modeled (thin solid lines and shadings; yellow for OLS, blue for MFR, and red for MeHiM) probability distributions of TC landfall intensity for (a) North Atlantic coastline (green, blue, and red segments shown on the map), (b) Northeastern USA from Maine to Virginia (green segment), (c) South-East USA from North Carolina to Florida plus Gulf Coast of the United States (blue segment), and (d) Gulf Coast of Mexico (red segment). The $15^{th}$-$85^{th}$ percentile uncertainty bonds of the simulations (shadings) are estimated from 100 realizations.
50

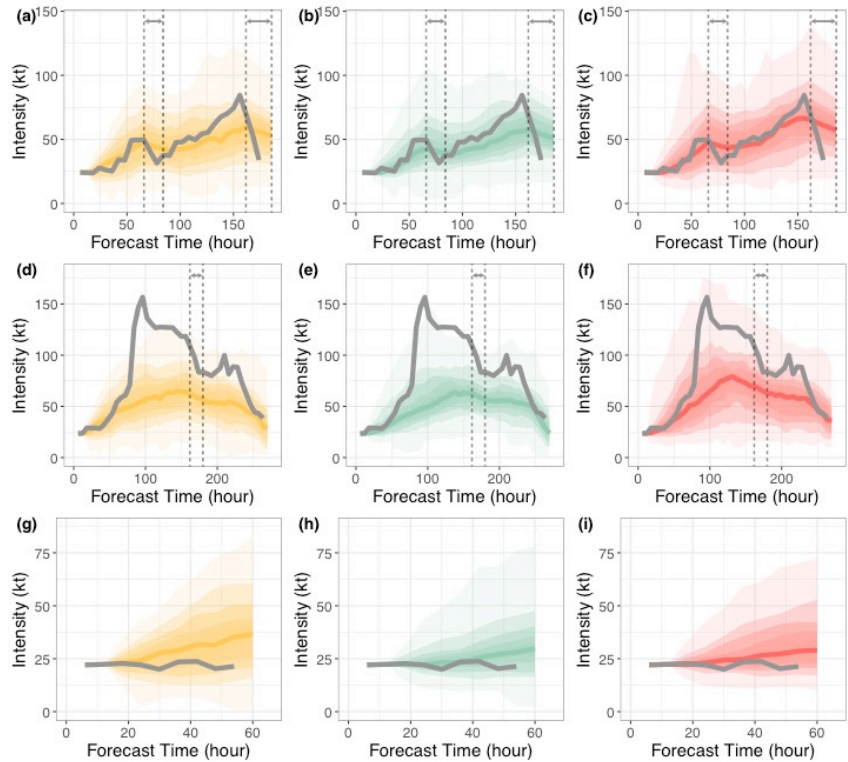

Fig. 10. Comparisons of observed (grey curve) and simulated intensity evolution for (a-c) Hurricane Alex (2010), (d-f) Hurricane Wilma (2005), and (h-i) an unnamed Tropical Depression in 2003. The simulated intensity evolutions from OLS (yellow) are shown in (a), (d), and (g); simulations from FMR (blue) are shown in (b), (d), and (h); and simulations from MeHiM (red) are shown in (c), (f), and (i). The colored shading boundaries represent the deciles of 100 realizations while the solid colored curve is the mean over the 100 realizations. The vertical dashed lines highlight the period of observations that are over land, during which the land model is applied. Over-land observations that are less than two steps (12 hours) are not highlighted.

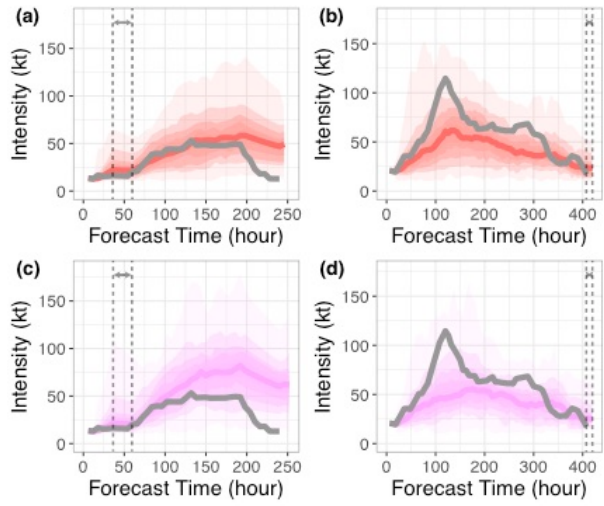

Fig. 11. Comparisons of MeHiM simulated intensity evolution with (red) and without (purple) ocean parameter. The observation is shown by the grey curve. Selected storms are (a, c) Tropical Storm Larry (2003) and (b, d) Hurricane Felix (1995). The vertical dashed lines highlight the period of observations that are over land, during which the land model is applied.

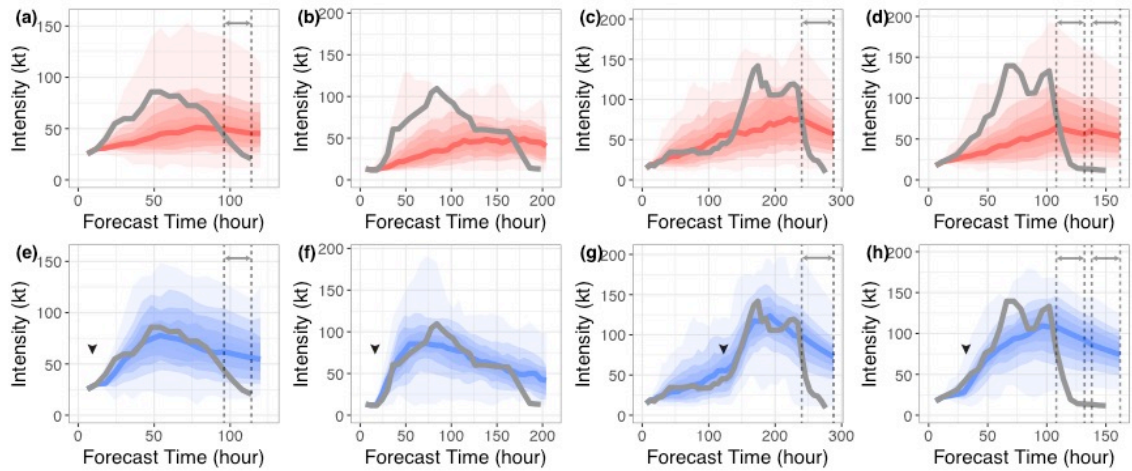

Fig. 12. Comparison of MeHiM simulated intensity evolution without (red) and with (blue) RI state correction. The observation is shown by the grey curve. Selected four RI storms are (a, e) Paula (2010); (b, f) Harvey (1981); (c, g) Andrew (1992); and (d, h) Felix (2007), from left to right. The black arrow indicates the occurrence of RI for each storm. The vertical dashed lines highlight the period of observations that are over land, during which the land model is applied.

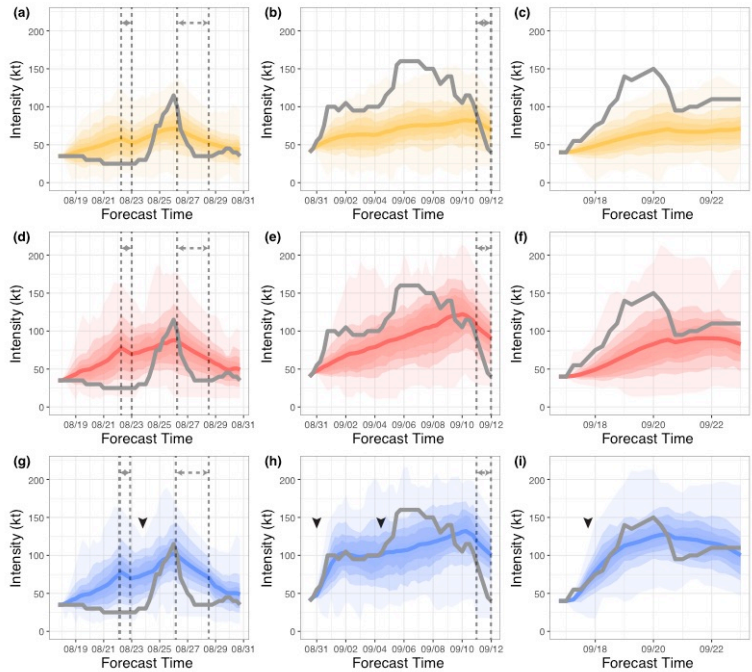

Fig. 13. Comparisons of observed (grey lines) and simulated intensity for (a, d, g) Hurricane Harvey (2017); (b, e, h) Hurricane Irma (2017); and (c, f, i) Hurricane Maria (2017). (a-c) are simulation results from OLS (yellow). (d-f) are simulations from MeHiM (red), and (g-i) are from MeHiM with RI state correction (blue). The black arrow indicates the occurrence of RI for each storm. The vertical dashed lines highlight the period of observations that are over land, during which the land model is applied.